\documentclass[12pt, draftclsnofoot, onecolumn]{IEEEtran}

\usepackage{amssymb}
\usepackage{amsmath}
\usepackage{cite}
\usepackage{url}
\usepackage{xcolor}
\usepackage{cite,graphicx,amsmath,amssymb}
\usepackage{subfigure}
\usepackage{fancyhdr}
\usepackage{mdwmath}
\usepackage{mdwtab}
\usepackage{caption}
\usepackage{amsthm}
\usepackage{setspace}
\usepackage{algorithm}
\usepackage{algorithmic}
\usepackage{float}

\newtheorem{remark}{Remark}
\newtheorem{theorem}{Theorem}

\newtheorem{lemma}{Lemma}

\newtheorem{corollary}{Corollary}

\begin{document}

\title{NOMA Empowered Integrated Sensing and Communication}
\author{

    Zhaolin~Wang,~\IEEEmembership{Graduate Student Member,~IEEE,}
    Yuanwei~Liu,~\IEEEmembership{Senior Member,~IEEE,}
    Xidong~Mu,~\IEEEmembership{Graduate Student Member,~IEEE,}
    Zhiguo Ding,~\IEEEmembership{Fellow,~IEEE,}
    and Octavia A. Dobre,~\IEEEmembership{Fellow,~IEEE} 

\thanks{Zhaolin Wang and Yuanwei Liu are with the School of Electronic Engineering
and Computer Science, Queen Mary University of London, London E1 4NS,
U.K. (e-mail: zhaolin.wang@qmul.ac.uk; yuanwei.liu@qmul.ac.uk).}
\thanks{Xidong Mu is with the School of Artificial Intelligence, Beijing University of Posts and Telecommunications, Beijing 100876, China (e-mail:
muxidong@bupt.edu.cn).}
\thanks{Zhiguo Ding is with the School of Electrical and Electronic Engineering, 
The University of Manchester, Manchester M13 9PL, U.K. (e-mail: zhiguo.ding@manchester.ac.uk).}
\thanks{Octavia A. Dobre is with the Faculty of Engineering and Applied Science, Memorial University, St. John's, NL A1B 3X5, Canada (e-mail:
odobre@mun.ca)}
}

\maketitle

\begin{abstract}
    A non-orthogonal multiple access (NOMA) empowered integrated sensing and communication (ISAC) framework is investigated. 
    A dual-functional base station serves multiple communication users employing NOMA, while the superimposed NOMA communication signal 
    is simultaneously exploited for target sensing. A beamforming design problem is formulated to maximize the weighted sum of the communication throughput and the 
    effective sensing power. To solve this problem, an efficient double-layer penalty-based algorithm is proposed by invoking 
    successive convex approximation. Numerical results show that the proposed NOMA-ISAC outperforms the conventional ISAC in the underloaded regime experiencing highly correlated channels and in the overloaded regime.
\end{abstract}

\section{Introduction}
\IEEEPARstart{R}{ecently}, with the global commercial deployment of the fifth-generation (5G) wireless communication networks, the concepts of 
beyond 5G (B5G) and six-generation (6G) wireless communication have received heated discussion. In contrast to 5G, it is envisioned that the 
networks in B5G and 6G will be designed to carry out simultaneously sensing and communication \cite{letaief2019roadmap}. 
Toward this trend, integrated sensing and communication (ISAC) \cite{liu2021integrated}, in which the radar sensing and wireless communication are integrated to share the same spectrum and infrastructure,
is proposed and have received growing attention from both academia \cite{zhang2021enabling} and industries \cite{tan2021integrated}.

Motivated by the advantages of ISAC like sharing spectrum and reducing cost, there has been a number of research contributions recently \cite{liu2018mu,liu2020beamforming,dong2020low}.
Nevertheless, we notice that these works do not consider the impact of spatially correlated channels and
overloaded regime on the ISAC system, which is more likely to encounter in the future wireless communication networks due to the massive number of connected devices \cite{liu2021evolution}. 
For the conventional multi-antenna technique employed in \cite{liu2018mu, liu2020beamforming, dong2020low}, the communication users will suffer from severe
inter-user interference when the channels are highly correlated or the system is overloaded because of the limited spatial degrees of freedom (DoFs). Thus, even the requirement of the communication-only system cannot be satisfied. 
In this case, it is generally impossible to integrate the sensing function into the communication system.
As a remedy, non-orthogonal multiple access (NOMA) can multiplex communication users in the power domain and mitigate the inter-user interference 
by exploiting successive interference cancellation (SIC) \cite{liu2017non}, which provides extra DoFs. Furthermore, NOMA can benefit ISAC by allowing more users to be served than the conventional multiple access techniques,
thus achieving higher spectral efficiency.
However, to the best of the authors' knowledge, the application of NOMA in ISAC has not been studied yet, which motivates this work.

In this letter, we propose a NOMA-ISAC framework, in which the transmitted superimposed signal is exploited for communication and 
sensing simultaneously and SIC is exploited for inter-user interference mitigation. We formulate a beamforming design problem for the maximization of the weighted sum of the communication throughput and the effective sensing power, 
subject to the minimum communication rate and radar-specific constraints. To solve this non-convex problem, 
we propose a double-layer penalty-based algorithm based on successive convex approximation (SCA). Our numerical results show that when the spatial correlation is high or 
the system is overloaded, the proposed NOMA-ISAC achieves better performance trade-off than the conventional ISAC.

\section{System Model and Problem Formulation}

\subsection{System Model}
A NOMA-ISAC system is proposed, which consists of a dual-functional base station (BS) equipped with an $N$-antennas uniform linear array (ULA),
$K$ single-antenna users indexed by $\mathcal{K} = \{1,\dots,K\}$, and $M$ radar targets indexed by $\mathcal{M} = \{1,\dots,M\}$. 

\subsubsection{Communication Model}
Different from existing research contributions \cite{liu2018mu, liu2020beamforming, dong2020low}, in this work, NOMA is employed at the BS for serving multiple communication users \cite{liu2017non}. 
Specifically, the BS transmits the superimposed signals of 
$\mathbf{w}_i s_i$ for $\forall i \in \mathcal{K}$ to all users, where $\mathbf{w}_i \in \mathbb{C}^{N \times 1}$ are beamformers
for delivering the information symbol $s_i$ to user $i$.
Therefore, the received signal $y_k$ at user $k$ is given by
\begin{equation}\label{eqn:received_signal}
    y_k = \mathbf{h}_k^H \sum_{i \in \mathcal{K}} \mathbf{w}_i s_i + n_k = \sum_{i \in \mathcal{K}} \mathbf{h}_k^H \mathbf{w}_i s_i + n_k,
\end{equation} 

\begin{figure} [t!]
    \centering
    \includegraphics[width=0.5\textwidth]{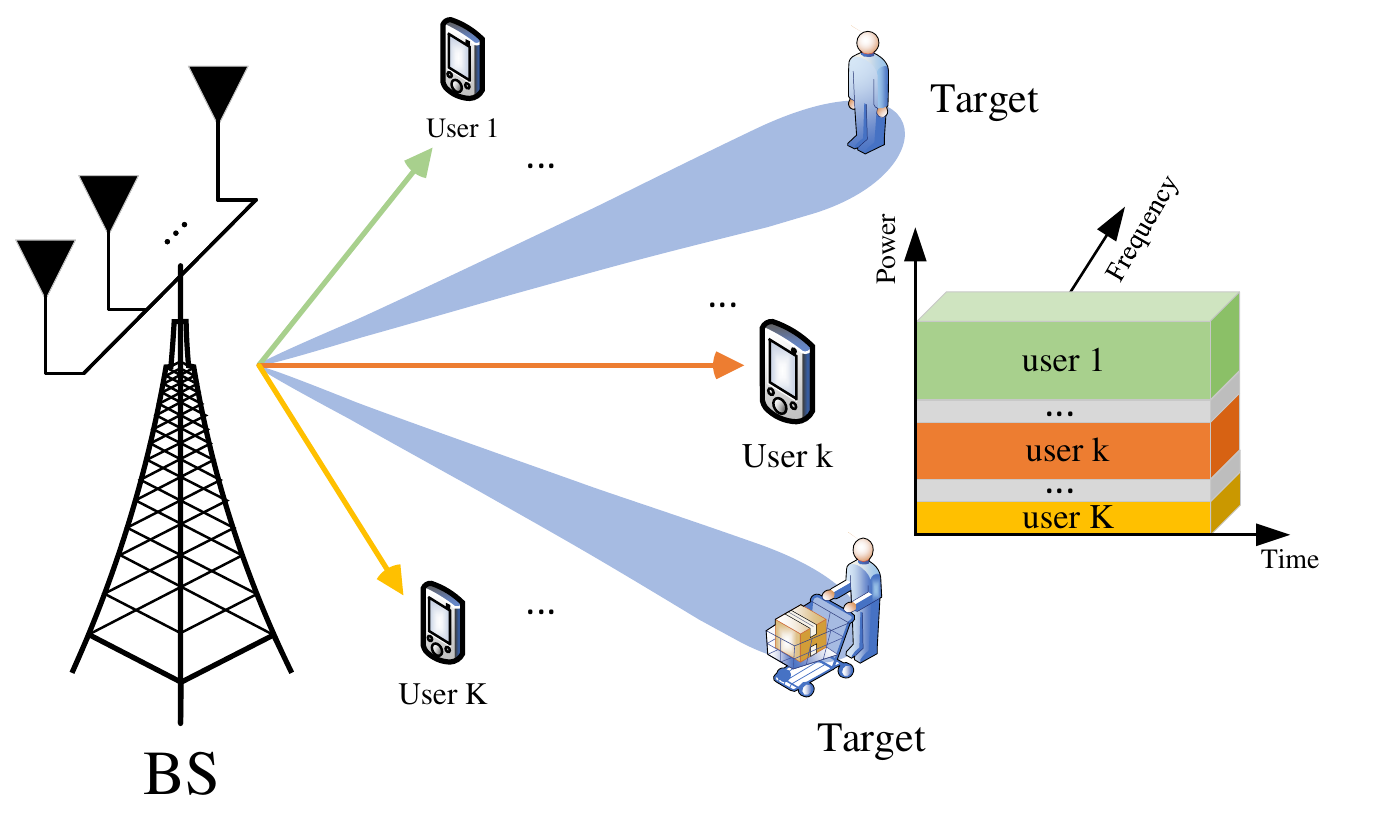}
     \caption{Illustration of the NOMA-empowered ISAC system.}
     \label{fig:system_model}
 \end{figure}
where $\mathbf{h}_k = \Lambda_k^{-1/2} \widetilde{\mathbf{h}}_k, \forall k \in \mathcal{K}$ denotes the BS-user channel, $\Lambda_k^{-1/2}$ and 
$\widetilde{\mathbf{h}}_k \in \mathbb{C}^{N \times 1}$ denote the large and small scale fading, respectively, and $n_k$ denotes the circularly symmetric complex Gaussian noise with variance $\sigma_n^2$. 
We assume that the users' indexes are in an increasing order with respect to their large-scale channel strength, 
i.e., $\Lambda_1^{-1} \le \Lambda_2^{-1} \le \dots \le \Lambda_K^{-1}$. Thus, user $1$ is the weakest user while user $K$ is the strongest user.
In NOMA, user $k$ first detects and removes the interference from all the weaker $j < k$ users by exploiting SIC, while treating the interference from all the stronger 
users $j > k$ as noise. Thus, the achievable rate of $s_k$ after SIC at user $k$ for $ \forall k \in \mathcal{K}, k \neq K$ is 
\begin{equation} \label{eqn:rate}
    R_{k \rightarrow k}  = \log_2 \left( 1 + \frac{|\mathbf{h}_k^H \mathbf{w}_k|^2}{ \sum_{i \in \mathcal{K}, i>k} |\mathbf{h}_k^H \mathbf{w}_i|^2 + \sigma_n^2  } \right).
\end{equation}
However, the symbol $s_k$ for user $k$ also need to be decodable at user $j$, for $j>k$ and $ \forall k \in \mathcal{K}, k \neq K$, to carry out SIC, yielding the following achievable rate
\begin{equation} \label{eqn:rate_SIC}
    R_{k \rightarrow j}  = \log_2 \left( 1 + \frac{|\mathbf{h}_j^H \mathbf{w}_k|^2}{ \sum_{i \in \mathcal{K}, i>k} |\mathbf{h}_j^H \mathbf{w}_i|^2 + \sigma_n^2  } \right).
\end{equation} 
Thus, the overall achievable rate of $s_k$ for $\forall k \in \mathcal{K}, k \neq K$ is
\begin{equation}\label{eqn:achievable_rate_overall}
    R_k = \min \{ R_{k \rightarrow k},\dots, R_{k \rightarrow K} \}.
\end{equation}  
At user $K$, the interference from all the other users is eliminated by SIC. Therefore, its achievable rate is given as 
\begin{equation}
    \label{eqn:rate_interference_free}
    R_K = \log_2 \left( 1 + \frac{|\mathbf{h}_K^H \mathbf{w}_K|^2}{\sigma_n^2}\right).
\end{equation}

Therefore, the communication throughput of the $K$ users is given by $R = \sum_{k \in \mathcal{K}} R_k$.

\subsubsection{Sensing Model}
In the ISAC system, the communication waveforms can be exploited to perform radar target sensing, but need to satisfy the sensing 
requirements, which is equivalent to design the covariance matrix of the transmitted signal \cite{stoica2007probing}. The covariance matrix is given by 
\begin{equation}\label{eqn:transmit_covariance_matrix}
    \mathbf{R}_{\mathbf{w}} =  \sum_{i \in \mathcal{K}} \mathbf{w}_i \mathbf{w}_i^H.
\end{equation}
With the prior information of target, the objective for sensing system is to maximize the effective sensing power, i.e., the power of probing signal in target directions \cite{stoica2007probing}, which is given as
\begin{equation}\label{eqn:transmit_beampattern}
    P(\theta_m) = \mathbf{a}^H(\theta_m) \mathbf{R}_{\mathbf{w}} \mathbf{a}(\theta_m),
\end{equation}
where $\theta_m, \forall m \in \mathcal{M}$ are target directions and $ \mathbf{a}(\theta_m) = [1, e^{j\frac{2\pi}{\lambda}d\sin({\theta_m})},...,e^{j\frac{2\pi}{\lambda}d(N-1)\sin({\theta_m})}]^T $ is the steering vector, where $\lambda$ and $d$ denote the carrier wavelength and antenna spacing, respectively.  
We assume that similar levels of sensing power is desired in the different target directions such that each target can be fairly tracked.
Furthermore, the cross-correlation between transmitted signals at any two target directions $\theta_k$ and $\theta_p$ is expected to be low such that the sensing system can perform adaptive localization \cite{stoica2007probing}.
The cross-correlation is given by $C(\theta_k, \theta_p) = | \mathbf{a}^H(\theta_k) \mathbf{R}_{\mathbf{w}} \mathbf{a}(\theta_p)|, \forall k \neq p \in \mathcal{M}$. 
The mean-squared cross-correlation of the $\frac{M^2-M}{2}$ pairs of sensing targets is given by 
\begin{equation}
    \bar{C} =  \frac{2}{M^2-M} \sum_{k=1}^{M-1} \sum_{p=k+1}^{M} C(\theta_k, \theta_p)^2.
\end{equation}
\begin{remark} \label{remark_1}
    \emph{In contrast to the conventional ISAC \cite{liu2018mu, liu2020beamforming, liu2018toward}, where the inter-user interference is merely mitigated by exploiting spatial 
    DoFs, the NOMA-ISAC system further employs SIC for mitigating inter-user interference, 
    see \eqref{eqn:rate}, \eqref{eqn:rate_SIC}, and \eqref{eqn:rate_interference_free}. Therefore, when the available spatial DoFs are limited (e.g., the underloaded regime with highly 
    correlated channels and overloaded regime), NOMA provides extra DoFs to guarantee the communication performance, 
    which enables the feasibility of integrating the sensing function. This will be verified via the numerical results in 
    Section \ref{sec:results}.}
\end{remark}

\subsection{Problem Formulation}
Given our NOMA-ISAC framework, we aim to maximize the weighted sum of communication throughput and effective sensing power, while satisfying 
the minimum communication rate of each user and radar-specific requirements. The resultant optimization problem is formulated as 
\begin{subequations} \label{problem:max_rate_power}
  \begin{align}
    \label{obj:max_rate_power}
    \max_{\mathbf{w}_k} \quad & \rho_c \sum_{k \in \mathcal{K}}R_k + \rho_r \sum_{m \in \mathcal{M}}P(\theta_m) \\
    \label{constraint:min_rate}
    \mathrm{s.t.} \quad & R_k \ge R_{\text{min},k}, \forall k \in \mathcal{K}\\
    \label{constraint:power_difference}
    & |P(\theta_k) - P(\theta_p)| \le P_{\mathrm{diff}}, \forall k \neq p \in \mathcal{M}\\
    \label{constraint:constant_modulus}
    &\mathrm{diag}\left(\sum_{i \in \mathcal{K}} \mathbf{w}_i \mathbf{w}_i^H \right) = \frac{P_t \mathbf{1}^{N \times 1}}{N},\\
    \label{constraint:cross_correlation}
    &\bar{C} \le \xi,
  \end{align}
\end{subequations}
where $\rho_c \ge 0$ and $\rho_r \ge 0$ are the regularization parameters; by varying them we can obtain the performance trade-off between communication and radar sensing.  
Here, \eqref{constraint:min_rate} guarantees the minimum rate of each user and \eqref{constraint:power_difference} ensures the similar levels of sensing power in different target directions.
The constraint \eqref{constraint:constant_modulus} is the constant per antenna constraint \cite{stoica2007probing}, where $P_t$ denotes the total transmit power.
Finally, the constraint \eqref{constraint:cross_correlation} ensures a desired upper bound of the mean-squared cross-correlation.
However, it is quite challenging to obtain the globally optimal solution for problem \eqref{problem:max_rate_power} due to the following reasons. On the one hand, the expression of achievable rate is neither convex nor concave,
which makes the objective function non-concave and the constraint \eqref{constraint:min_rate} non-convex. On the other hand, the quadratic form of the covariance matrix makes the constraints
\eqref{constraint:power_difference} and \eqref{constraint:constant_modulus} non-convex. In the following, we propose an efficient iterative algorithm to 
obtain a suboptimal solution by invoking SCA \cite{sun2016majorization}. 

\section{Proposed Solution}

In this section, we develop an SCA-based double-layer iterative algorithm. Firstly, we define 
$\mathbf{W}_k \triangleq \mathbf{w}_k \mathbf{w}_k^H$, which satisfies $\mathbf{W}_k \succeq 0$, $\mathbf{W}_k = \mathbf{W}_k^H$,
and $\mathrm{rank}(\mathbf{W}_k)=1$. Then, the problem \eqref{problem:max_rate_power} can be reformulated as 
\begin{subequations} \label{problem:max_rate_power_2}
  \begin{align}
    \max_{ \gamma_k, \mathbf{W}_k}  & f(\rho_c, \rho_r, \gamma_k, \mathbf{W}_k) = \rho_c \sum_{k \in \mathcal{K}} \gamma_k + \rho_r \! \!\! \sum_{m \in \mathcal{M}} P(\theta_m) \\
    \label{constraint:max_min_simple}
    \mathrm{s.t.} \quad & R_k \ge \gamma_k, \forall k \in \mathcal{K},\\
    \label{constraint:min_rate_2}
    &\gamma_k \ge R_{\text{min},k}, \forall k \in \mathcal{K} \\
    \label{constraint:semidefinite}
    & \mathbf{W}_k \succeq 0, \mathbf{W}_k = \mathbf{W}_k^H, \forall k \in \mathcal{K} \\
    \label{constraint:rank}
    & \mathrm{rank}(\mathbf{W}_k) = 1, \forall k \in \mathcal{K}\\
    & \eqref{constraint:power_difference}-\eqref{constraint:cross_correlation}.
  \end{align}
\end{subequations}
Furthermore, We define ${\bf H}_k \triangleq \mathbf{h}_k \mathbf{h}_k^H$. Then, for $j \ge k$ and $k \in \mathcal{K}, k \neq K$, the constraint \eqref{constraint:max_min_simple} can be rewritten as 
\begin{align} \label{eqn:rate_constraint}
  R_{k \rightarrow j} = \log_2\Big(\sigma_n^2 + \sum_{i \in \mathcal{K}, i \ge k} \mathrm{Tr}\left( \mathbf{H}_j \mathbf{W}_i \right) \Big)
  \underbrace{- \log_2\Big(\sigma_n^2 + \sum_{i \in \mathcal{K}, i > k} \mathrm{Tr}\left( \mathbf{H}_j \mathbf{W}_i \right) \Big)}_{F_{j,k}} \ge \gamma_k. 
\end{align}
The non-convexity of this constraint lies in the second term $F_{j,k}$. 
To address this, we invoke the SCA. By using the first-order Taylor expansion at point $\big(\mathbf{W}_1^n,\dots,\mathbf{W}_K^n)$, we have 
\begin{align} \label{eqn:taylor_expansion_rate}
  F_{j,k} \ge \widehat{F}_{j,k} \triangleq -\log_2 \Big( \sigma_n^2 + \sum_{i \in \mathcal{K}, i > k} \mathrm{Tr}\left( \mathbf{H}_j \mathbf{W}_i^n \right) \Big)
  - \frac{\sum_{i \in \mathcal{K}, i > k} \mathrm{Tr}\big( \mathbf{H}_j \left( \mathbf{W}_i - \mathbf{W}_i^n \right) \big) }{\left( \sigma_n^2 + \sum_{i \in \mathcal{K}, i > k} \mathrm{Tr}\left( \mathbf{H}_j \mathbf{W}_i^n \right) \right)\ln 2}.
\end{align}
Then, we define 
\vspace{-0.2cm}
\begin{equation}
  \widehat{R}_{k \rightarrow j} \triangleq \log_2\Big(\sigma_n^2 + \sum_{i \in \mathcal{K}, i \ge k} \mathrm{Tr}\left( \mathbf{H}_j \mathbf{W}_i \right) \Big) + \widehat{F}_{j,k},
\end{equation}
which is a lower bound of $R_{k \rightarrow j}$. By exploiting it, the constraint \eqref{constraint:max_min_simple} can be approximated by $\widehat{R}_{k \rightarrow j} \ge \gamma_k$.  
Thus, problem \eqref{problem:max_rate_power_2} can be reformulated as
\begin{subequations} \label{problem:max_rate_power_3}
    \begin{align}
      \max_{ \gamma_k, \mathbf{W}_k} \quad &f(\rho_c, \rho_r, \gamma_k, \mathbf{W}_k) \\
      \label{constraint:max_min_1}
      \mathrm{s.t.} \quad & \widehat{R}_{k \rightarrow j} \ge \gamma_k, j \ge k, \forall k \in \mathcal{K}, k \neq K,\\
      \label{constraint:max_min_2}
      & R_K \ge \gamma_K,\\
      & \eqref{constraint:power_difference} - \eqref{constraint:cross_correlation}, \eqref{constraint:min_rate_2} - \eqref{constraint:rank}.
    \end{align}
\end{subequations}
For this optimization problem, the non-convexity is only from the rank-one constraint \eqref{constraint:rank}. 
Generally, the semidefinite relaxation (SDR) \cite{luo2010semidefinite} is exploited to solve this problem by omitting the rank-one constraint. Then, the eigenvalues decomposition
or Gaussian randomization is used to reconstruct the rank-one solution from the general-rank solution obtained by SDR, which may lead 
to a significant performance loss and not ensure the feasibility of the reconstructed matrix. To avoid these drawbacks, we attempt to 
transform the rank-one constraints to a penalty term in the objective function \cite{mu2021noma}, which can also be solved by SCA.
Toward this idea, we firstly introduce an equivalent equality constraint:
\begin{equation} \label{eqn:rank_1_equality}
    \| \mathbf{W}_k \|_* - \| \mathbf{W}_k \|_2 = 0, k \in \mathcal{K},
\end{equation} 
where $\| \cdot \|_*$ is the nuclear norm, which is the sum of singular values of the matrix, and $\| \cdot \|_2$ is the spectral norm,
which is the largest singular values of the matrix. Thus, when the matrix $\mathbf{W}_k$ is a rank-one matrix, the equality \eqref{eqn:rank_1_equality} holds.
Otherwise, as $\mathbf{W}_k$ is semidefinite, we must have that the sum of singular values is larger than the largest singular value, i.e., $\| \mathbf{W}_k \|_* - \| \mathbf{W}_k \|_2 > 0$. 
In order to obtain a rank-one matrix, we introduce a penalty term to the objective function based on \eqref{eqn:rank_1_equality}, yielding
\begin{subequations} \label{problem:max_rate_power_4}
    \begin{align}
      \max_{ \gamma_k, \mathbf{W}_k}  &f(\rho_c, \rho_r, \gamma_k, \mathbf{W}_k) - \frac{1}{\eta} \sum_{k \in \mathcal{K}} \left( \| \mathbf{W}_k \|_* - \| \mathbf{W}_k \|_2 \right)\\
      \mathrm{s.t.} & \quad \eqref{constraint:power_difference} - \eqref{constraint:cross_correlation}, \eqref{constraint:min_rate_2}, \eqref{constraint:semidefinite}, \eqref{constraint:max_min_1}, \eqref{constraint:max_min_2}.
    \end{align}
\end{subequations}
However, the second term in the penalty term makes the objective not convex. By exploiting the first-order 
Taylor expansion at point $\mathbf{W}_k^n$, its upper bound of is given by
\begin{align} \label{eqn:taylor_expansion}
  -\| \mathbf{W}_k \|_2 \le \widehat{\mathbf{W}}_k^n
  \triangleq -\| \mathbf{W}_k^n \|_2 - \mathrm{Tr} \big[ \mathbf{v}_{\max,k}^n (\mathbf{v}_{\max,k}^n)^H \left( \mathbf{W}_k - \mathbf{W}_k^n \right) \big], 
\end{align}
where $\mathbf{v}_{\max,k}^n$ is the eigenvector corresponding to the largest eigenvalue of $\mathbf{W}_k^n$. Thus, the problem \eqref{problem:max_rate_power_4}
can be approximated by the following problem 
\begin{subequations} \label{problem:max_rate_power_5}
  \begin{align}
    \max_{ \gamma_k, \mathbf{W}_k} \quad &f(\rho_c, \rho_r, \gamma_k, \mathbf{W}_k) - \frac{1}{\eta} \sum_{k \in \mathcal{K}} \left( \| \mathbf{W}_k \|_* + \widehat{\mathbf{W}}_k^n \right)\\
    \mathrm{s.t.} \quad & \eqref{constraint:power_difference} - \eqref{constraint:cross_correlation}, \eqref{constraint:min_rate_2}, \eqref{constraint:semidefinite}, \eqref{constraint:max_min_1}, \eqref{constraint:max_min_2}.
  \end{align}
\end{subequations}
The problem \eqref{problem:max_rate_power_5} is a quadratic semidefinite program (QSDP), which can be efficiently solved by the CVX toolbox \cite{cvx}.

\begin{algorithm}[htb]
  \caption{Proposed double-layer penalty-based algorithm for solving problem \eqref{problem:max_rate_power}.}
  \label{alg:A}
  \begin{algorithmic}[1]
      \STATE{Initialize the feasible $\mathbf{W}_k^0, \forall k \in \mathcal{K}$.}
      \REPEAT
      \STATE{$n \gets 0$.}
        \REPEAT
          \STATE{Update $\mathbf{W}_k^{n+1}$ by solving \eqref{problem:max_rate_power_5} with $\mathbf{W}_k^n$, $\forall k \in \mathcal{K}$.}
          \STATE{$n \gets n+1$.}
        \UNTIL{the fractional reduction of the objective function value falls below a predefined threshold $\varepsilon_1$. }
      \STATE{$\mathbf{W}_k^0 \gets \mathbf{W}_k^n, \forall k \in \mathcal{K}$.}
      \STATE{$\eta \gets \epsilon \eta$.}
      \UNTIL{$\sum_{k \in \mathcal{K}}\left(\| \mathbf{W}_k^n \|_* - \| \mathbf{W}_k^n \|_2\right) \le \varepsilon_2$.}

  \end{algorithmic}
\end{algorithm}

It is worth noting that the choice of parameter $\eta$ plays an important role in the objective function. If this parameter is chosen 
to be $\eta \rightarrow 0$ ($\frac{1}{\eta} \rightarrow \infty$), the rank of matrix $\mathbf{W}_k$ will be definitely one. Nevertheless, 
in this case, we cannot obtain a good solution regarding the maximization of throughput and effective sensing power, since the objective function
is dominated by the penalty term. To tackle this, we can initialize a large $\eta$ to obtain a good starting point for the throughput and the effective sensing power.
Then, by gradually reducing $\eta$ to a sufficiently small value via $\eta = \epsilon \eta, 0<\epsilon<1$, an overall suboptimal solution can be obtained. 
This procedure is terminated when the penalty term is sufficiently small, i.e., $\sum_{k \in \mathcal{K}}\left(\| \mathbf{W}_k \|_* - \| \mathbf{W}_k \|_2\right) \le \varepsilon_2$.
The overall algorithm to problem \eqref{problem:max_rate_power} is summarized in Algorithm \ref{alg:A}. The complexity of this algorithm is $\mathcal{O}(I_o I_i (K^{6.5}N^{6.5} \log(1/e)))$, where 
$I_o$ and $I_i$ are the number of iterations of the outer and inner layers, $e$ is the solution accuracy, and $\mathcal{O}(K^{6.5}N^{6.5} \log(1/e))$ is the complexity for solving the QSDP \eqref{problem:max_rate_power_5} \cite{liu2020beamforming}.  

\section{Numerical Results} \label{sec:results}

\begin{figure} [t!]
    \centering
    \includegraphics[width=0.47\textwidth]{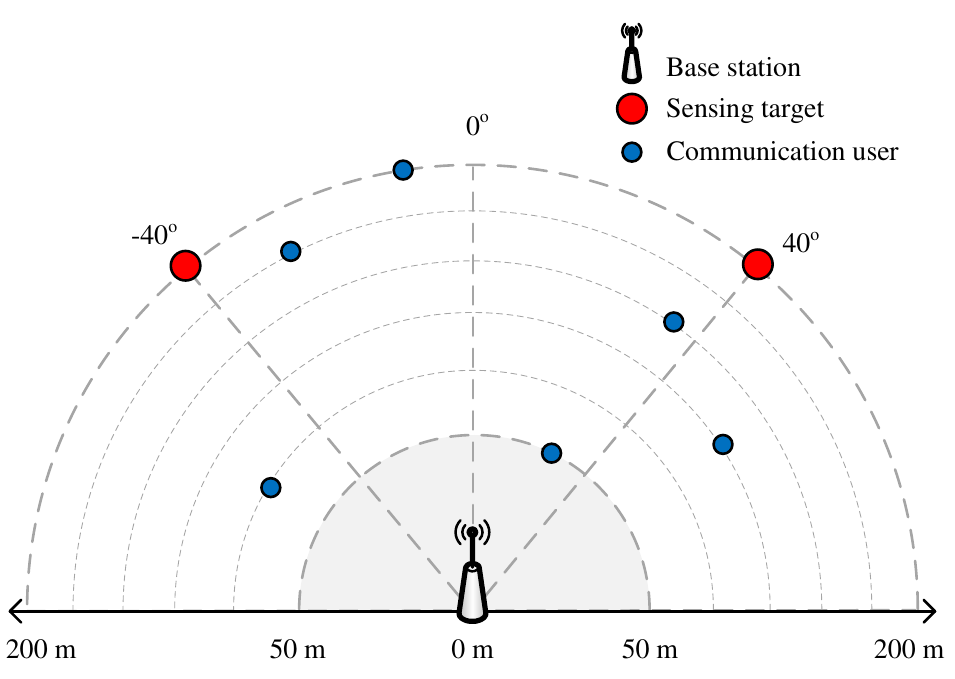}
    \caption{Simulation setup.}
    \label{fig:simulation_setup}
    \vspace{-0.6cm}
\end{figure}

In this section, the numerical results are provided to demonstrate the characteristics of NOMA in ISAC systems. 
As shown in Fig. \ref{fig:simulation_setup}, we assume a BS equipped with a ULA with $N=4$ antennas, serving $K=2$ or $6$ communication users and tracking $M=2$ radar targets in $\theta_1=-40^\circ$ and $\theta_2=40^\circ$.
The overall power budget is $P_t=20$ dBm and the noise power at users is $\sigma_n^2=-120$ dBm. The channels between BS and users are 
assumed to experience Rayleigh fading with the path loss of $\Lambda_k (\mathrm{dB}) = 32.6 + 36.7 \log_{10} (d_k)$. 
In particular, the path loss model is defined based on the 3GPP propagation environment \cite[Table B.1.2.1-1]{3gpp.36.814}.
The fading model follows \cite{kermoal2002stochastic}
\begin{equation}
    \widetilde{\mathbf{H}} = \mathbf{H}_w \mathbf{R}_{\widetilde{\mathbf{H}}}^{1/2},
\end{equation}
where $\widetilde{\mathbf{H}} = \big[ \mathbf{h}_1/\|\mathbf{h}_1\|,\dots,\mathbf{h}_K/\|\mathbf{h}_K\| \big]$ and 
$\mathbf{H}_w$ is the normalized Rayleigh fading matrix satisfying $\mathbb{E}\left[ \mathbf{H}_w^H \mathbf{H}_w \right] = \mathbf{I}$. 
The matrix $\mathbf{R}_{\widetilde{\mathbf{H}}}$ is the spatial correlation matrix of $\widetilde{\mathbf{H}}$. Its $(i, j)$-th entry indicates the spatial correlation 
between the channels of user $i$ and user $j$, the norm of which is $t^{|i-j|}$ for $t \in [0,1]$. The users are equally spaced
between the distances of $50$ m and $200$ m from BS. We set $R_{\min}=1$ bit/s/Hz, $P_{\mathrm{diff}}=10$, and $\xi=10$. The initial penalty factor of Algorithm \ref{alg:A} is set to $\eta=10^5$.
Finally, the convergence thresholds of inner and outer layers are set to $\varepsilon_1 = 10^{-2}$ and $ \varepsilon_2 = 10^{-4}$. 

\subsection{Convergence of Algorithm \ref{alg:A}}
\begin{figure}[t!]
  \centering
  \includegraphics[width=0.47\textwidth]{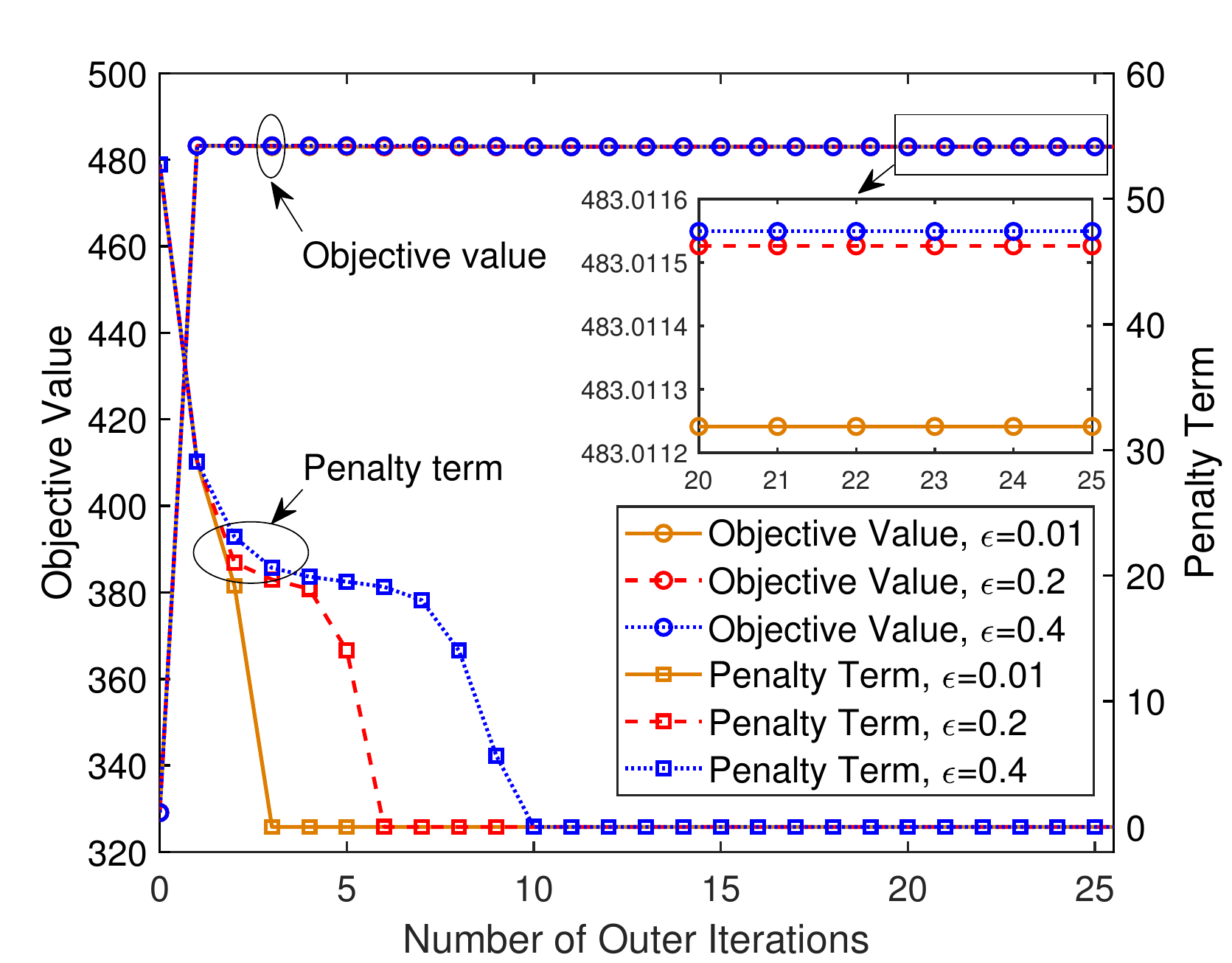}
  \caption{Convergence of Algorithm \ref{alg:A}.}
  \label{fig:convergence}
\end{figure}

In Fig. \ref{fig:convergence}, the convergence behavior of the proposed
algorithm over one random channel realization is studied with $K=6$, $t=0$, $\rho_c=10$, and $\rho_r=1$.  
We can observe that for any values of the reduction factor $\epsilon$, the objective value quickly converges to a stable value while the penalty term converges to almost zero after several outer iterations.
It reveals that the proposed algorithm is capable of finding a feasible rank-one solution with high performance. Furthermore, it can be seen that as the value of $\epsilon$ 
becomes smaller, the proposed algorithm has higher convergence speed while achieves lower objective value, i.e., worse system performance, which is a trade-off.
Thus, in the following simulation, we set $\epsilon=0.2$, which achieves the suitable convergence speed and system performance simultaneously. 

\subsection{Baseline}
For comparison, we consider the conventional ISAC system without the employment of NOMA \cite{liu2018mu}, where the achievable rate at user $k$ is given by 
\begin{equation}
    \label{eqn:sdma_sinr}
    R_k^{\mathrm{b}} = \log_2 \left( 1 + \frac{|\mathbf{h}_k^H \mathbf{w}_k|^2}{\sum_{i \in \mathcal{K}, i \neq k}| \mathbf{h}_k^H \mathbf{w}_i|^2 + \sigma_n^2} \right).
\end{equation} 
The corresponding problem of maximizing the throughput $R^{\mathrm{b}} = \sum_{k \in \mathcal{K}} R_k^{\mathrm{b}}$ and effective sensing power at the sensing targets can be solved using Algorithm \ref{alg:A}
with the interference term in \eqref{eqn:sdma_sinr}.

\subsection{Performance Trade-off}

\begin{figure}[t!]
    \centering
    \subfigure[Underloaded $N=4, K=2$]{
        \includegraphics[width=0.47\textwidth]{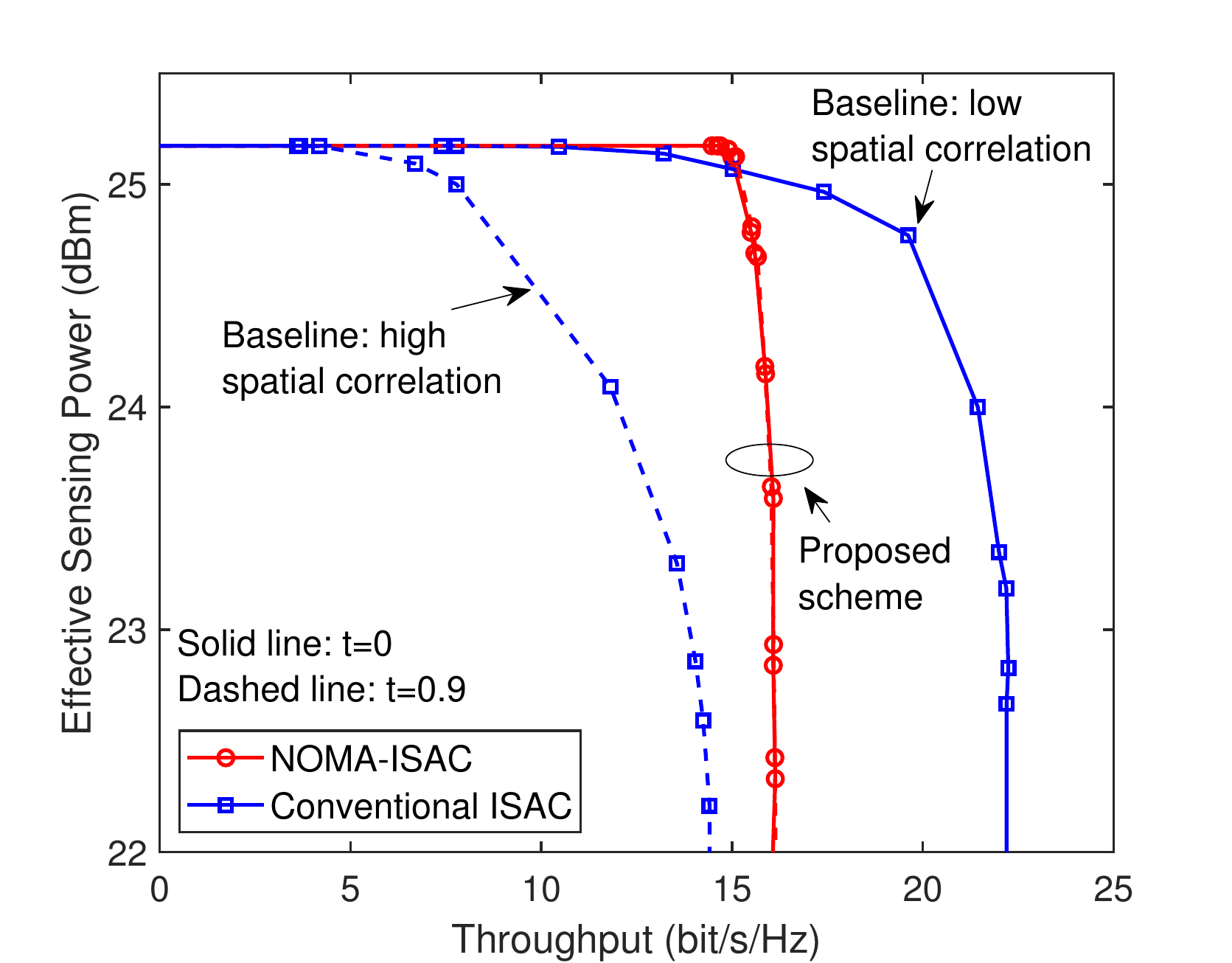}
        \label{fig:underloaded}
    }
    \subfigure[Overloaded $N=4, K=6$]{
        \includegraphics[width=0.47\textwidth]{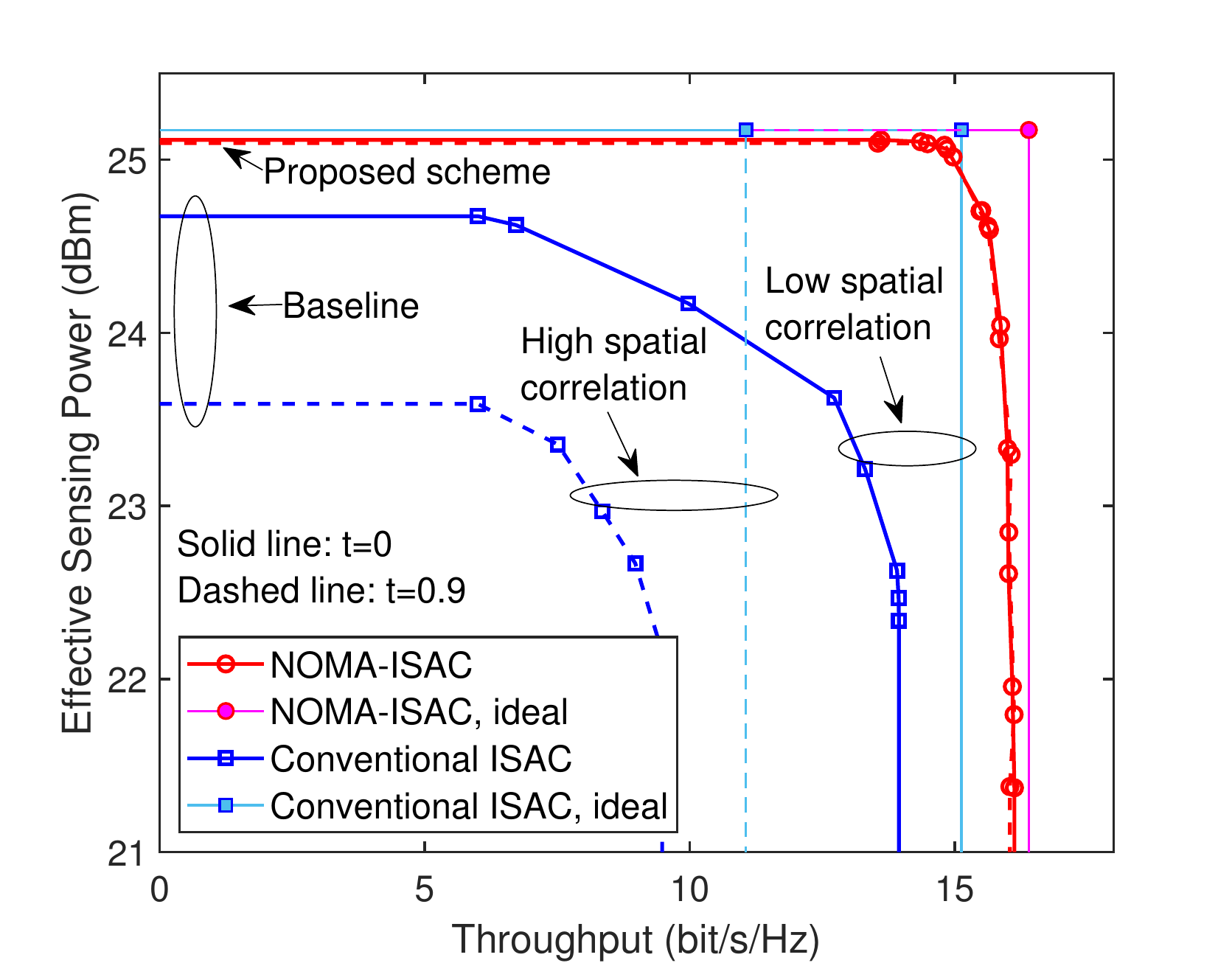}
        \label{fig:overloaded}
    }

     \caption{Trade-off between throughput and effective sensing power.}
    \label{fig:trade-off}
\end{figure}

In Fig. \ref{fig:trade-off}, we demonstrate the performance trade-off, i.e., the communication throughput versus the effective radar sensing power, of NOMA-ISAC and the conventional ISAC.
The results are obtained via Monte Carlo simulation over $400$ random channel realizations. Two cases are considered, namely the underloaded regime ($K=2$) and the overloaded regime ($K=6$).
As seen in Fig. \ref{fig:trade-off}, in both underloaded and overloaded regimes, the spatial factor has no effect on the NOMA-ISAC system. The reason is that the communication throughput is dominated by the strongest NOMA user.
However, the performance of the conventional ISAC system is subject to the spatial factor, i.e., as the spatial correlation increases, the performance achievable area becomes smaller.
Specifically, in the underloaded regime (Fig. \ref{fig:underloaded}), the NOMA-ISAC outperforms the conventional ISAC when the spatial correlation is high,
but the result is opposite when the spatial correlation is low. 
In the overloaded regime (Fig. \ref{fig:overloaded}), compared to the conventional ISAC, the NOMA-ISAC achieves considerable gain, which becomes even greater when the spatial correlation is high.
This is because the inter-user interference cannot be well mitigated in the conventional ISAC system due to the limited spatial DoFs in the overloaded regime.
Then, more resources are needed by the conventional ISAC to meet the communication requirements, thus leading to the limited sensing performance.
However, for the proposed NOMA-ISAC, although the system is overloaded, the inter-user interference can still be mitigated by SIC, which provides more DoFs to be exploited for the radar sensing.
The above results underscore the importance of employing NOMA in the ISAC system when the communication system is overloaded or the channels are highly spatially correlated  
and verify \textbf{Remark \ref{remark_1}}. 

Furthermore, we also demonstrate the ideal ISAC system in Fig. \ref{fig:overloaded}, i.e., communication and sensing systems work independently, with no effect on each other. 
The communication throughput and effective sensing power in the ideal ISAC is obtained by removing the communication and sensing from problem \eqref{problem:max_rate_power}, respectively.
For the conventional ISAC system, it can be seen that there is a significant gap between the real case and the ideal case. However, for the NOMA-ISAC system, there is only a slight performance gap and 
the performance upper bounds of communication and radar sensing can be nearly achieved simultaneously.

\subsection{Transmit Beampattern}

\begin{figure}[t!]
    \centering
    \subfigure[Underloaded $N=4, K=2, t=0$]{
        \includegraphics[width=0.47\textwidth]{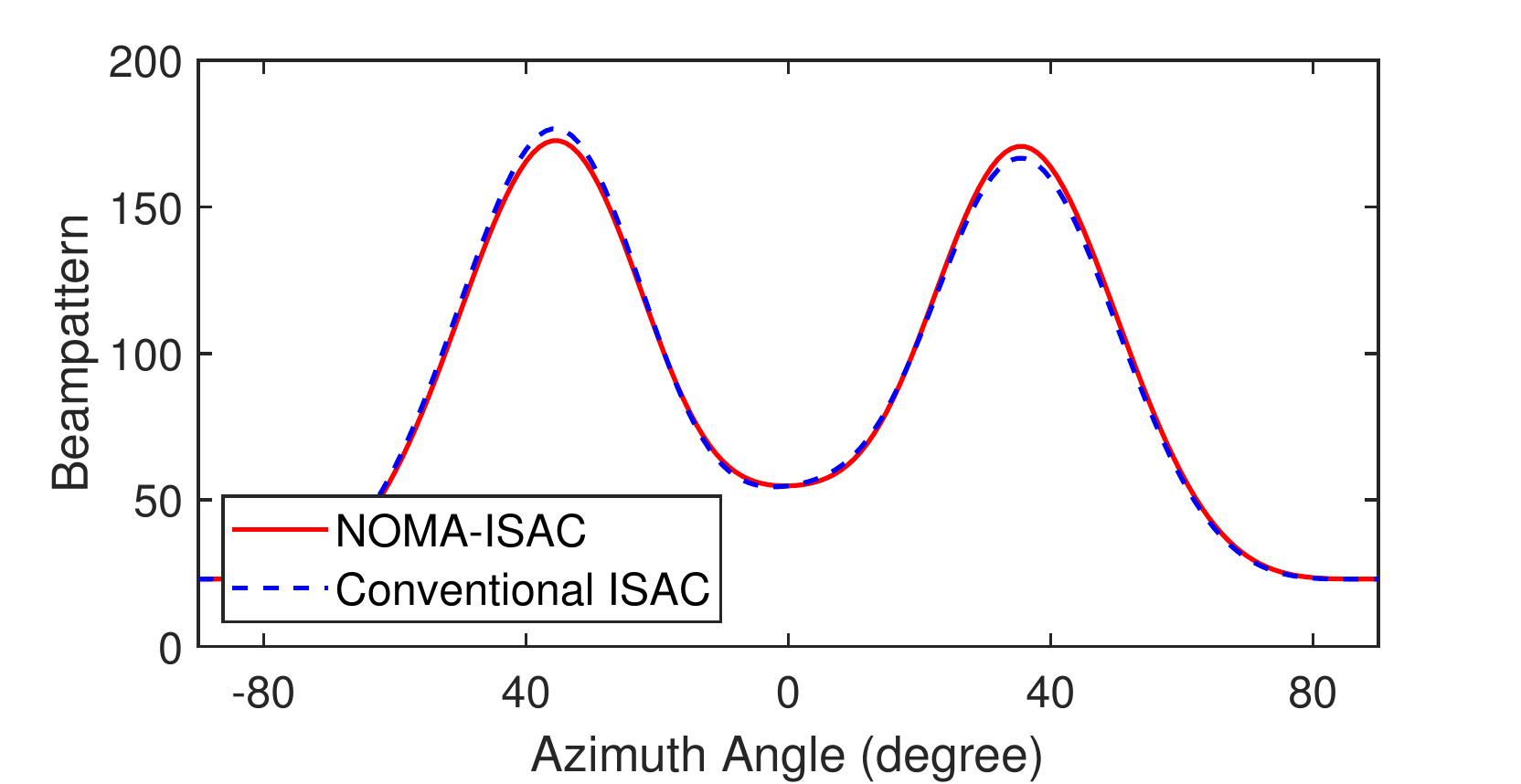}
        \label{fig:underloaded_patterm}
    }
    \subfigure[Overloaded $N=4, K=6, t=0$]{
        \includegraphics[width=0.47\textwidth]{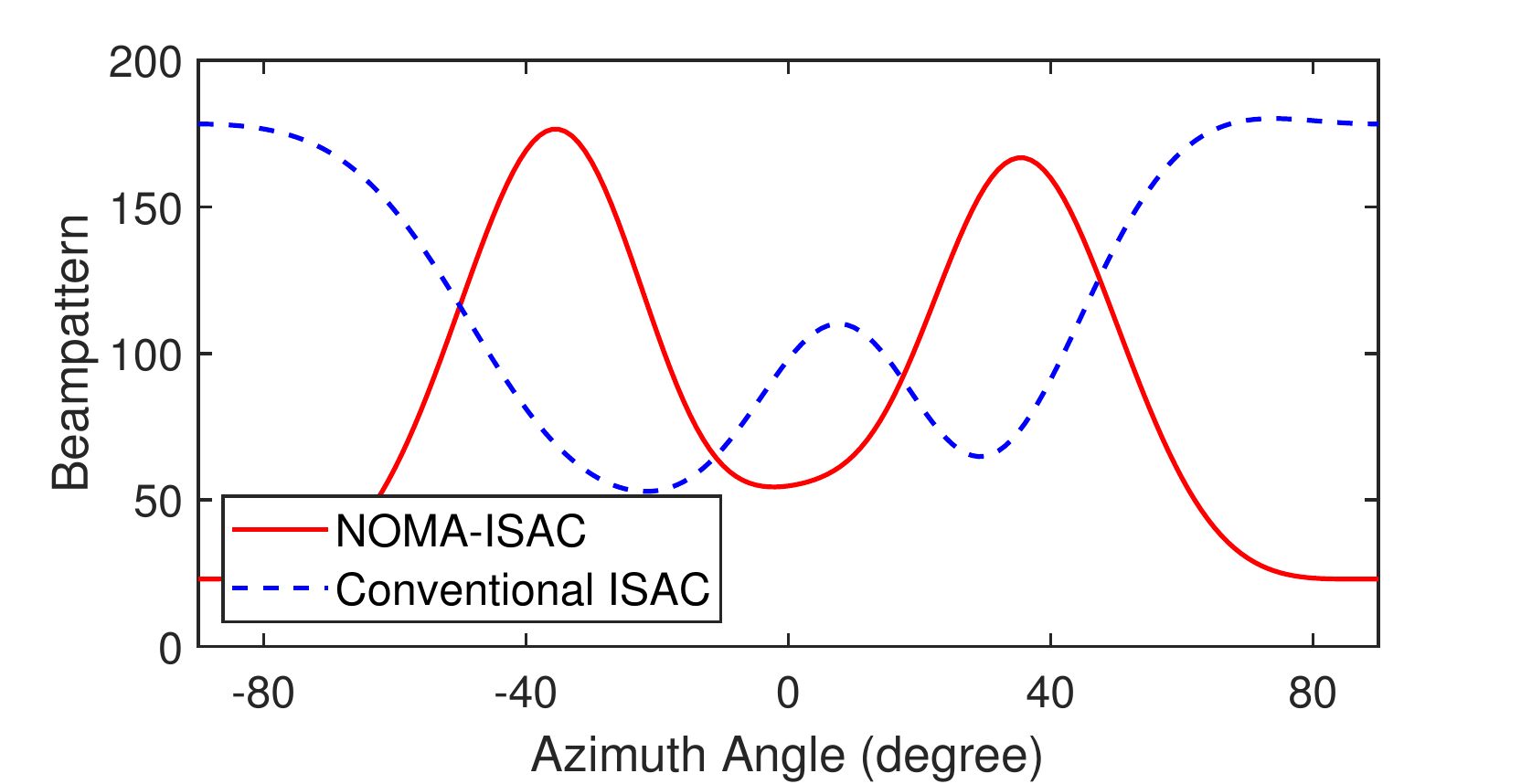}
        \label{fig:overloaded_patterm}
    }
    \caption{Obtained transmit beampattern by different schemes when the communication throughput is $13.5$ bit/s/Hz.}
    \label{fig:beampatterm}
\end{figure}

In Fig. \ref{fig:beampatterm}, we present the obtained transmit beampattern by the proposed NOMA-ISAC and the conventional ISAC over one random channel realization when the communication throughput 
is $13.5$ bit/s/Hz. We set the spatial correlation factor as $t=0$. Similarly, both underloaded ($K=2$) and overloaded regimes ($K=6$) are considered. It can be observed that in the underloaded 
regime, both NOMA-ISAC and the conventional ISAC can achieve the dominant peak of the transmit beampattern in the directions of interest, i.e., $-40^\circ$ and $40^\circ$. In the overloaded regime, the
dominant peaks can still be achieved by the proposed NOMA-ISAC, while the conventional ISAC experiences severe power leakage in the undesired directions, leading to the significant sensing performance degradation.
These results further emphasize the importance of NOMA in terms of guaranteeing the sensing performance when the ISAC system is overloaded and also verify \textbf{Remark \ref{remark_1}.}

\section{Conclusion}

A NOMA empowered ISAC system has been proposed.
A tailor-made beamforming optimization problem was formulated to maximize the weighted sum of communication throughput and effective sensing power 
subject to the minimum communication rate of each user and the radar-specific requirement.
To solve the non-convex problem, a double-layer penalty-based algorithm was developed to obtain a suboptimal solution. Our numerical results indicated that 
the NOMA-ISAC system achieves a better communication-sensing trade-off than the conventional ISAC system when the system is overloaded or the channel spatial correlation is high. 
Furthermore, in the overloaded regime, the performance of the NOMA-ISAC system is close to the ideal ISAC system, which means that in practice, 
it can provide high quality communication and radar sensing functions simultaneously. 

\bibliographystyle{IEEEtran}
\bibliography{mybib}

\end{document}